\documentstyle[prl,twocolumn,aps,epsfig]{revtex}
\begin{document}
\draft
\title{Phase Conjugation of Trapped Bose-Einstein Condensates}
\author{Elena V. Goldstein and Pierre Meystre}
\address{Optical Sciences Center, University of Arizona, Tucson, AZ 85721\\
June 11, 1998 \\ \medskip}
\author{\small\parbox{14.2cm}{\small\hspace*{3mm}
We consider a multicomponent atomic Bose-Einstein condensate optically
trapped in a far-off resonant dipole trap. Drawing an analogy with the optical
situation, we show that this system can be regarded as an matter-wave
analog of optical multiwave mixing. We concentrate specifically on
condenstates in the hyperfine ground state $F=1$, in which case a simple
analogy with optical four-wave mixing can be established. This opens
up the way to realize matter-wave phase conjugation, whereby an
atomic beam can be ``time-reversed.''  In addition to transfering 
population between a "central" mode and incident and
retroreflecting beams, matter-wave phase conjugation 
also offers novel diagnostic tools to study the
coherence properties of condensates, as well as to measure the relative
scattering lengths of hyperfine sublevels.
\\[3pt]PACS numbers: 03.75.-b 03.75.Fi 05.30.Jp 42.65.Hw}}
\maketitle
\narrowtext

\section{Introduction}
The experimental observation of Bose-Einstein condensation in low density
atomic vapors \cite{BECexp} has triggered
a flurry of theoretical activity \cite{BECtheor}.
Theoretical predictions on the condensate dynamics, ground state population
and spectrum of elementary excitations have been made and are in excellent
agreement with experiments.

At the same time, further experimental advances have led to the
realization of multicomponent condensates, both in $^{87}$Rb and in
$^{23}$Na. In the first case, sympathetic cooling, together with a fortuitous
coincidence in the scattering lengths of the spin states $m=-1/2$ and $m=-3/2$
led to the coexistence of both components in a magnetic trap
\cite{MyaBurGhr97}. A multicomponent
condensate was also achieved with the three hyperfine ground state
components of sodium in a far-off resonant dipole trap \cite{StaAndChi98}.
These results, along with further experiments involving 
condensates in double well
potentials \cite{MewAndKur97}, have now led to considerable theoretical
work on the static and dynamic properties of multicomponent
condensates, including studies of their true ground state
\cite{DouGro}, analyses of the elementary excitations spectrum,
and the determination of their instability regions
\cite{DouExc}.
In addition, multicomponent condensates open up the way to novel schemes
to launch vortices and permanent currents
\cite{Vort} and to the study of novel
phenomena such as vector and quadrupole spin wave modes, the
topological and energetical instabilities of doubly quantized singular
vortices \cite{Ho98}, etc. The two-body interactions
characteristic of spin-1 condensates can lead to intercomponent
coupling via two-body collisions.

In the zero-temperature limit, a $q$-component Bose Einstein condensate can
be thought of as a $q$-mode system, whereby the various modes are coupled
by two-body (and possibly higher-order) collisions which result in the
exchange of particles between these modes. As such, they correspond to
a situation quite similar to that of multiwave mixing in nonlinear optics.
A main difference is of course that in the case of matter waves, the
coupling is due to the collisions, which find their origin in the
electromagnetic interaction between the atoms. Collisions can then be
thought of as the effective atom-atom interaction
resulting from the elimination of the electromagnetic
field from the system dynamics. This is to be contrasted with
the optical case, where multimode mixing relies on
the interaction of the electromagnetic field with a common atomic sample
whose dynamics is traced over. In that case, it is the elimination
of the material dynamics that results in an effective field-field coupling.
This observation is the origin of nonlinear atom optics, which
is the matter-wave equivalent of nonlinear optics.

A close analogy can easily be established between the 
dynamics of a spin-1 condensate as realized in
sodium experiments and the situation of degenerate
four-wave mixing in optics, as we demonstrate explicitly in this paper.
In particular, for situations where the $m=0$ state is macroscopically
populated while the $m=\pm 1$ states are weakly excited, one can think
of the first state as a ``pump'' or "central"
mode, while $m=\pm 1$ form side modes,
which are coupled via the pump, leading to the familiar effects of
degenerate four-wave mixing, including phase conjugation. This is the
effect that we study in detail in this work. We then show how matter-wave
phase conjugation can be used as a diagnostic tool to study
the coherence properties of Schr\"odinger fields, as well as the
relative scattering lengths of the states involved.

Matter-wave phase conjugation has previously been studied, but in a
situation where the coupling between the partial matter waves was
induced by the near-resonant electric dipole-dipole interaction
\cite{GolPlaMey96}. As such,
it relied explicitly on having a substantial population of electronically
excited atoms, and the incoherent effects of spontaneous emission rapidly
destroyed the coherent wave coupling responsible for phase conjugation.
In contrast, the situation with a condensate of dipole-trapped sodium atoms
does not suffer from this drawback: since we are considering ground-state
atoms in a far-off resonant trap with hyperfine levels coupled primarily via
ground state collisions, spontaneous emission is certainly negligible.
In addition, the fact that the atoms are in a trap changes the situation
somewhat from the free-space geometry considered in our earlier work, since
the atomic sample can easily be tightly confined in the transverse
dimensions and hence does not suffer from free-space diffraction. Our main
result is to demonstrate that a trapped condensate can then be used as a
phase-conjugate mirror for a weak atomic beam, thereby effectively
``time-reversing'' it.

Section II describes our physical model, and derives the coupled-wave
equations for the three components of the condensate in the Hartree regime.
This is applied in section III to the discussion of matter-wave phase
conjugation in two-dimensional atomic traps.  We concentrate explicitly
on the undepleted pump regime, and show how the phase conjugate signal
depends explicitly on the relative scattering lengths of the hyperfine
levels involved. Finally, the possible experimental verification of our
predictions, as well as a summary and outlook are given in section IV.

\section{Physical model}

We consider a condensate of $^{23}$Na atoms in their $F=1$ hyperfine ground
state, with three internal atomic states $|F=1,m=-1\rangle$,
$|F=1,m=0\rangle$ and $|F=1,m=1\rangle$ of degenerate energies in the absence
of magnetic fields. It is described by the three-component vector
Schr\"odinger field
\begin{equation}
\bbox{\Psi}({\bf r},t)
=\{\Psi_{-1}({\bf r},t),\Psi_{0}({\bf r},t),\Psi_{1}({\bf r},t)\}
\end{equation}
which satisfies the bosonic commutation relations
\begin{equation}
[\Psi_i({\bf r},t), \Psi_j^\dagger({\bf r}',t)]
=\delta_{ij}\delta({\bf r}-{\bf r}').
\end{equation}
Accounting for the possibility of two-body collisions, its dynamics
is described by the second-quantized Hamiltonian
\begin{eqnarray}
& &{\cal H}=\int d {\bf r} \bbox{\Psi}^\dagger({\bf r},t)H_0
\bbox{\Psi}({\bf r},t)
\nonumber\\
& &+\int \{d {\bf r}\}\bbox{\Psi}^\dagger({\bf r}_1,t)
\bbox{\Psi}^\dagger({\bf r_2},t)V({\bf r}_1-{\bf r}_2)
\bbox{\Psi}({\bf r}_2,t)\bbox{\Psi}({\bf r}_1,t),
\label{ham2}
\end{eqnarray}
where the single-particle Hamiltonian is
\begin{equation}
H_0={\bf p}^2/2M + V_{trap}
\end{equation}
and the trap potential is of the general form
\begin{equation}
V_{trap} = \sum_{m=-1}^{+1} U({\bf r})|F=1,m\rangle\langle F=1,m|.
\end{equation}
Here ${\bf p}$ is the center-of-mass momentum of the atoms of mass $M$ and
$U({\bf r})$, the  effective dipole trap potential for
atoms in the $|1,m \rangle$ hyperfine state, is independent of
$m$ for a non-magnetic trap.

The general form of the two-body interaction $V({\bf r}_1-{\bf r}_2)$
has been discussed in detail in Refs. \cite{Ho98,ZhaWal98}. We reproduce
its main features for the sake of clarity. Consider situations where
the hyperfine spin $F_i = 1$ of the individual atoms is preserved.
We label the hyperfine states of the combined system of hyperfine spin
${\bf F} = {\bf F}_1 + {\bf F}_2$ by $|f,m \rangle$ with $f = 0, 1,2$
and $m=-f,\ldots,f$.
In the shapeless approximation, it can then be shown that the two-body
interaction is of the general form \cite{Ho98}
\begin{equation}
 V({\bf r}_1-{\bf r}_2)=\delta({\bf r}_1-{\bf r}_2)\sum_{f=0}^{2}
\hbar g_f {\cal P}_f,
\label{pot}
\end{equation}
where
\begin{equation}
g_f=4\pi\hbar a_f/M,
\end{equation}
${\cal P}_f\equiv \sum_m|f,m\rangle\langle f,m|$ is the projection operator 
which projects the pair of atoms into a total hyperfine $f$ state
and $a_f$ is the $s$-wave scattering length for the channel of total 
hyperfine spin $f$. For bosonic atoms only even $f-$states contribute, so
that
\begin{eqnarray}
V({\bf r}_1-{\bf r}_2)&=&\hbar \delta({\bf r}_1-{\bf r}_2)(g_2{\cal P}_2+
g_0{\cal P}_0) \nonumber\\
&=&\frac{\hbar}{2}\delta({\bf r}_1-{\bf r}_2)
\left(c_0+c_2{\bf F}_1\cdot{\bf F}_2 \right).
\end{eqnarray}
In this expression,
\begin{eqnarray}
c_0 &=& 2(g_0+2g_2)/3 \nonumber \\
c_2 &=& 2(g_2-g_0)/3.
\end{eqnarray}
Substituting this form of $V({\bf r}_1-{\bf r}_2)$ into the second-quantized
Hamiltonian (\ref{ham2}) leads to
\begin{eqnarray}
& &{\cal H}=\sum_m\int d{\bf r} \Psi_m^\dagger({\bf r},t)
\left[\frac {{\bf p}^2}{2M}+U({\bf r})\right]\Psi_m({\bf r},t)\nonumber\\
& &
+\frac{\hbar}{2}\int d{\bf r}\{(c_0+c_2)
[\Psi_1^\dagger\Psi_1^\dagger\Psi_1\Psi_1
+\Psi_{-1}^\dagger\Psi_{-1}^\dagger\Psi_{-1}\Psi_{-1}
\nonumber\\
& &
+2\Psi_0^\dagger\Psi_0
(\Psi_1^\dagger\Psi_1+\Psi_{-1}^\dagger\Psi_{-1})]
+c_0\Psi_0^\dagger\Psi_0^\dagger\Psi_0\Psi_0
\nonumber\\
& &
+2(c_0-c_2)\Psi_1^\dagger\Psi_1
\Psi_{-1}^\dagger\Psi_{-1}
\nonumber\\
& &
+2c_2(\Psi_1^\dagger\Psi_{-1}^\dagger\Psi_0\Psi_0+H.c. )\}.
\label{ham22}
\end{eqnarray}

This form of the Hamiltonian is quite familiar in quantum optics, where it
describes four-wave mixing between a pump beam and two side-modes,
which are identified with the field operators $\Psi_0$ and $\Psi_{\pm 1}$
in the present situation. Specifically, we observe that the three terms
in the two-body Hamiltonian which are quartic in one of the field operators
only, i.e. of the form $\Psi_i{^\dagger} \Psi_i^{\dagger} \Psi_i \Psi_i$
can be readily interpreted as self-defocussing terms, corresponding to
the fact that the two-body potential is, for a positive scattering length and
a scalar field, analogous to a defocussing cubic nonlinearity in optics. The 
terms involving two ``modes'', i.e. of the type 
$\Psi_i^\dagger \Psi_i \Psi_j^\dagger \Psi_j$,
conserve the individual mode populations of the modes and simply lead to
phase shifts. Finally, the terms involving the central mode $\Psi_0$ and
{\em both} side-modes are the contributions of interest to us, since they
correspond to a redistribution of atoms between the ``pump'' mode
$\Psi_0$ and the side-modes $\Psi_{\pm 1}$, e.g. by annihilating two atoms
in the central mode and creating one atom each in the side-modes. This is
the kind of interaction that leads to phase conjugation in quantum optics,
except that in that case the modes in question are modes of the
Maxwell field instead of the Schr\"odinger field. Note also that a similar
mechanism is at the origin of amplification in the Collective
Atom Recoil Laser (CARL) \cite{BonSal94,MooMey982}, 
except that in that latter case, the Schr\"odinger field
mode coupling is induced by optical transitions in the atoms.

In the Hartree approximation, which is well justified for condensates
at $T=0$, the many-body problem reduces to an
effective single particle problem for the Hartree wave function
$\phi_m({\bf r},t)$. It is easily shown that to be governed by the
system of coupled nonlinear Schr\"odinger equations
\cite{NLAO}
\begin{eqnarray}
& &i\dot{\phi}_{-1}({\bf r},t)=\frac{1}{\hbar}
\left[\frac{{\bf p}^2}{2M}+U({\bf r})
\right] \phi_{-1}+N\{c_2\phi_0\phi_0\phi_1^\star
\nonumber\\
& &+[(c_0+c_2)(|\phi_{-1}|^2+|\phi_0|^2)
+(c_0-c_2)|\phi_{1}|^2]\phi_{-1}\}
\nonumber\\
& &i\dot{\phi}_{0}({\bf r},t)=\frac{1}{\hbar}
\left[\frac{{\bf p}^2}{2M}+U({\bf r})
\right] \phi_0+N\{c_0|\phi_0|^2\phi_0
\nonumber\\
& &+(c_0+c_2)(|\phi_{-1}|^2+|\phi_1|^2)\phi_{0}
+2c_2\phi_1\phi_{-1}\phi_0^\star\}
\nonumber\\
& &i\dot{\phi}_{1}({\bf r},t)=\frac{1}{\hbar}
\left[\frac{{\bf p}^2}{2M}+U({\bf r})
\right] \phi_1+N\{c_2\phi_0\phi_0\phi_{-1}^\star
\nonumber\\
& &+[(c_0+c_2)(|\phi_{1}|^2+|\phi_0|^2)
+(c_0-c_2)|\phi_{-1}|^2]\phi_{1}\}.
\label{schro}
\end{eqnarray}
Just as in the familiar quantum optics case, we consider in
the following a situation where the central mode, described by the
Hartree wave function $\phi_0$, is strongly populated initially, while
the side-modes $\phi_{\pm 1}$ are weakly populated. In other words, we
consider the phase conjugation of a weak atomic beam from a reasonably large
condensate. In that case, it is appropriate to introduce the matter-wave
optics equivalent of the undepleted pump approximation, whereby
\begin{equation}
{\dot {\phi}}_0 \simeq 0.
\label{undep}
\end{equation}
In that case, the problem reduces to a set of coupled mode equations for
the two side-modes $\phi_{\pm 1}$, the central mode acting as a catalyst
for the coupling between them.

\section{Phase conjugation in dipole traps}

In what follows we consider atomic samples confined in two-dimensional
harmonic trap.  The trap potential $U({\bf r})$ which is as we recall
independent of the atomic internal state $m$ \cite{Ho98} for a dipole trap,
is taken to be of the harmonic form
\begin{equation}
U({\bf r})=M\omega_0^2 (x^2+y^2)/2
\end{equation}
for simplicity. That is, we assume that the dipole trap confines the atoms
in the transverse plane $(x,y)$, but not in the longitudinal direction
$z$. This geometry allows one to consider side-modes propagating along
that axis, rather than bouncing back and forth in an elongated trap.
In case of tight confinement in the transverse direction, we can assume
to a good approximation that the transverse structure of the condensate
is not significantly altered by many-body interactions and is
determined as the ground-state solution of the transverse potential.

Expressing the Hartree wave function associated with the hyperfine level
$m$ as
\begin{equation}
\phi_m({\bf r}, t) = \varphi_\perp (x,y) \varphi_m(z,t) e^{-i\omega_0 t},
\label{anzats0}
\end{equation}
we then have
\begin{eqnarray}
\hbar\omega_0 \varphi_\perp(x,y)&=&\left [ -\frac{\hbar^2}{2M}
\left (\frac{\partial ^2}{\partial x^2}+\frac{\partial ^2}{\partial y^2}
\right)\right.
\nonumber\\
& &+\left.\frac {M\omega^2_0}{2}(x^2+y^2)\right]\varphi_\perp(x,y).
\end{eqnarray}
Substituting this expression into Eqs. (\ref{schro}) and projecting
out the transverse part of the wave function yields the coupled
one-dimensional Gross-Pitaevskii (coupled-mode) equations
\begin{eqnarray}
& &i\dot{\varphi}_{-1}(z,t)=-\frac{\hbar}{2M}\frac{\partial^2}{\partial z^2}
{\varphi}_{-1}+N\eta\{c_2\varphi_0\varphi_0\varphi_1^\star
\nonumber\\
& &+[(c_0+c_2)(|\varphi_{-1}|^2+|\varphi_0|^2)
+(c_0-c_2)|\varphi_{1}|^2]\varphi_{-1}\}
\nonumber\\
& &i\dot{\varphi}_{0}(z,t)=-\frac{\hbar}{2M}\frac{\partial^2}{\partial z^2}
{\varphi}_{0}+N\eta\{c_0|\varphi_0|^2\varphi_0
\nonumber\\
& &+(c_0+c_2)(|\varphi_{-1}|^2+|\varphi_1|^2)\varphi_{0}
+2c_2\varphi_1\varphi_{-1}\varphi_0^\star\}
\nonumber\\
& &i\dot{\varphi}_{1}(z,t)=-\frac{\hbar}{2M}\frac{\partial^2}{\partial z^2}
{\varphi}_{1}+N\eta\{c_2\varphi_0\varphi_0\varphi_{-1}^\star
\nonumber\\
& &+[(c_0+c_2)(|\varphi_{1}|^2+|\varphi_0|^2)
+(c_0-c_2)|\varphi_{-1}|^2]\varphi_{1}\},
\label{schro1}
\end{eqnarray}
where
\begin{equation}
\eta=\frac{\int dx dy |\varphi_\perp(x,y)|^4}{\int dx dy
|\varphi_\perp(x,y)|^2} .
\label{eff}
\end{equation}
The physical situation we have in mind is that of a weak ``probe'' in the
hyperfine state $m=-1$ propagating toward a large condensate in
state $m=0$ and at rest in the dipole trap, and generating a
backward-propagating conjugate matter wave in the hyperfine state $m = -1$
(see Fig. 1). Hence we express the longitudinal component of the Hartree
wave function as
$${\bbox{\varphi}}(z,t)\equiv\left(\begin{array}{c}
\varphi_{-1}(z,t) \\ 
\varphi_{0}(z,t) \\ 
\varphi_{1}(z,t)  \end{array}\right)
=\left(\begin{array}{c}\psi_{-1}(z,t) e^{-ikz} e^{-i\omega t} \\
2\psi_{0} \cos(kz) e^{-i\omega t} \\ \psi_{1}(z,t) e^{-ikz} e^{-i\omega t} 
\end{array}\right),
$$
\centerline{\psfig{figure=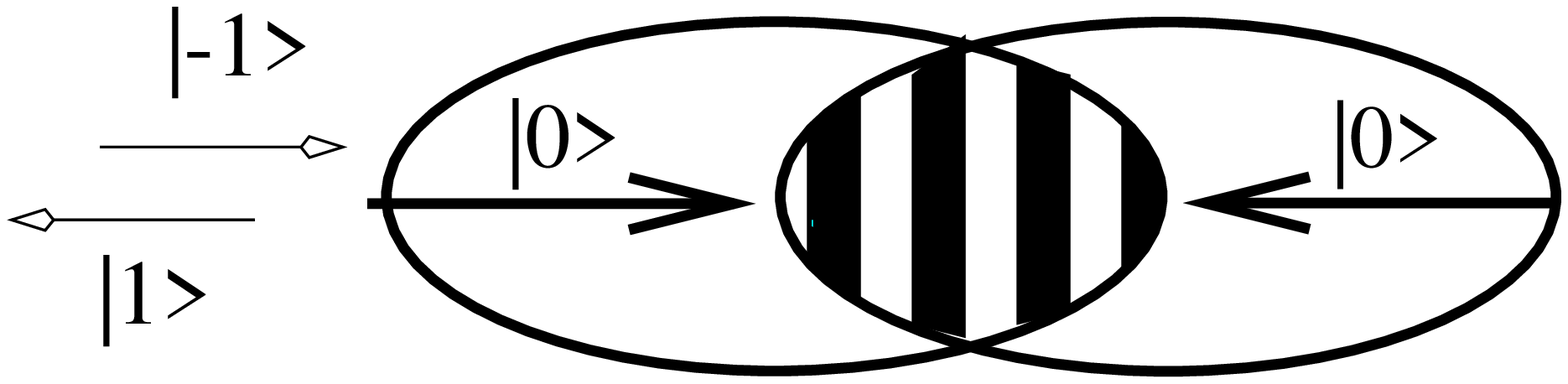,width=8.6cm,clip=}}
\begin{figure}
\caption{
Geometry of the matter wave phase conjugation with 3-component vector
fields.}
\end{figure}
\noindent where the slowly varying envelopes $\psi_m$ of the Hartree
wave function components $m = \pm 1$ satisfy the familiar inequalities
\begin{equation}
|\frac{\partial^2}{\partial z^2}\psi_m|\ll k|\frac{\partial}{\partial z}\psi_m|
\ll k^2 |\psi_m|
\end{equation}
and we have additionally invoked the undepleted pump approximation 
(\ref{undep}).
Note that in this ansatz the ``pump'' wave function $\varphi_0$ is described
by a standing wave. This spatial structure is required in order to achieve
momentum conservation, a direct consequence of the fact that a standing
wave can be viewed as a superposition of two counterpropagating atomic
waves. A state with such a periodic spatial structure can be achieved
for instance by interfering two condensates \cite{AndTowMie97},
in a grating matter-wave
interferometer, \cite{OvcKozDen98} or in CARL \cite{MooMey982}.
To the first order in the probe and signal fields, this geometry
leads to a linearized system of two coupled-mode equations for the probe
and condensate fields. In the stationary state they reduce to
\begin{eqnarray}
i\frac{\hbar k}{2M}\frac{\partial}{\partial z}\psi_{-1}(z)&=&-N\eta
[2(c_0+c_2)\rho_0\psi_{-1}(z)
\nonumber\\
&+&c_2\psi_{0}^2 \psi_{1}^\star(z)] ,
\nonumber\\
i\frac{\hbar k}{2M}\frac{\partial}{\partial z}\psi_{1}^\star(z)&=&-N\eta
[2(c_0+c_2)\rho_0\psi_1^\star(z)
\nonumber\\
&+&c_2\psi_{0}^2 \psi_{-1}(z)] ,
\label{phcon}
\end{eqnarray}
where $\rho_0 = |\psi_0|^2$.

The form of these equations is familiar from optical
phase conjugation and their solution is well-known. Before giving them
explicitly, though, we note that they contain two contributions. For
instance, the equation for the phase conjugate wave $\psi_1^\star$ contains
a term proportional to the density $\rho_0$ of the condensate  and the
field itself. In the absence of the second term, it would simply lead
to a phase shift of $\psi_1^\star$. Physically, it results from the
self-interaction of the conjugate field, catalyzed by the condensate
(pump) component. Its origin can be traced back to the
term proportional to $\Psi_1^\dagger\Psi_{1}^\dagger\Psi_0\Psi_0$ in
the Hamiltonian (\ref{ham22}).  The second term, in contrast, couples the
two side-modes via the condensate and is responsible for phase conjugation.
Note that it is not proportional to the condensate density $\rho_0$, but
rather to $\psi_0^2.$ We return to this point later on.

The general solution of Eqs. (\ref{phcon}) reads \cite{Fis83}
\begin{eqnarray}
\psi_{-1}(z)&=&\frac{e^{i\alpha z}}{\cos(|\kappa| L)}
\left(\right.- ie^{-i\beta}\sin(|\kappa|z)\psi_{1}^\star(L)
\nonumber\\
&+&\left.\cos(|\kappa|(z-L))\psi_{-1}(0) \right)
\nonumber\\
\psi_{1}(z) &=& \frac{e^{i\alpha z}}{\cos(|\kappa| L)}\left(
\right. \cos(|\kappa|z) \psi_1(L)
\nonumber\\
&+&i \left.e^{-i\beta}\sin(|\kappa|(z-L))\psi_{-1}^\star(0)\right),
\end{eqnarray}
where
\begin{equation}
\alpha=2N\eta (c_0+c_2)\rho_0,
\end{equation}
\begin{equation}
\kappa=\frac{N\eta c_2\psi_0^2}{\hbar k/2M} ,
\label{kappa}
\end{equation}
and
\begin{equation}
e^{i\beta}=\kappa/|\kappa|.
\end{equation}

For the probe $\psi_{-1}(0)$ incident at $z=0$ and no incoming conjugate
signal $\psi_1(L)=0$, the conjugate wave in the input plane $z=0$ becomes
\begin{equation}
\psi_{1}(0)=-ie^{-i\beta}\tan(|\kappa|L)\psi_{-1}^\star(0),
\end{equation}
which demonstrates that the interaction of the probe and the
condensate results in the generation of a counterpropagating phase-conjugated
signal. Note that the intensity of the conjugate wave exceeds 
that of the incoming
wave for $\pi/4<|\kappa|L<3\pi/4$ and phase conjugation oscillations (PCO)
\cite{Fis83} can occur for $|\kappa|L=\pi/2$, the so-called oscillation
condition.

\section{Experimental feasibility and outlook}

In order to determine the feasibility of matter-wave phase conjugation in
state of the art experiments, we briefly discuss the values of
the oscillation parameter $|\kappa|L$ that can be achieved in current 
$^{23}$Na BEC experiments.

From the definition (\ref{kappa}) we have
\begin{equation}
|\kappa|L \sim N\eta \frac{a_2-a_0}{3k} ,
\end{equation}
where we have taken that due to normalization
$\psi_{0}^2 L\simeq \rho_0 L\sim 1$ and that \cite{Ho98}
\begin{equation}
c_2=4\pi\hbar(a_2-a_0)/3M
\end{equation}
with $a_0$ and $a_2$ being the singlet and triplet state scattering lengths
respectively. For sodium, these scattering lengths are estimated as \cite{Ho98}
$(a_2-a_0)/3\sim 0.04 a_2\sim 10^{-10}$m. In the MIT optical
confinement experiments \cite{StaAndChi98} the number of trapped atoms is
of the order $5-10\cdot10^6$ 
and the transverse dipole trap frequency $\omega_0$ is
of the order of $10^4$ sec$^{-1}$, so that the transverse ground state size
of the condensate $a_\perp\equiv \sqrt{\hbar/m\omega_\perp}\sim 0.25\mu$m.
From Eq.(\ref{eff}) we have $\eta\sim a_\perp^{-2} \sim10^{13}$m$^{-2}$. As a
result the oscillation parameter is $|\kappa| L\sim 10^{10}/k$ where $k$ 
is as we recall the wave number of a Schr\"odinger field.
In case the condensate sidemodes are obtained by diffraction
on a standing light wave \cite{OvcKozDen98}, 
we have $k=2\pi/\lambda\sim 10^{7}$m$^{-1}$ and
thus $|\kappa| L\sim 10^{3}$. This means that the
oscillation condition  $|\kappa|L=\pi(2n+1)/2$ (n - integer)
can be met in current BEC experiments.

In addition to its interest from a nonlinear atom optics point of view,
matter-wave phase conjugation could also be used as a diagnostic tool
for Bose-Einstein condensates. For instance, we noted that the parameter
$|\kappa|L$ is proportional to the difference in scattering lengths between
the singlet and triplet states. Hence, this quantity could in principle
be inferred from phase conjugation measurements. In addition, we recall
that the phase conjugate signal is not determined by the condensate
density $\rho_0$, but rather by $\psi_0^2$. While the distinction between
the two is expected to be minimal for large condensates, and is essentially
ignored in the Hartree and undepleted pump approach of the present paper, 
this will no longer be the case for smaller condensates. In such situations,
phase conjugation provides one with a probe of the coherence properties
of the condensate. Future work will analyze these aspects of the problem,
as well as the role of higher-order correlation functions in the atom
statistics of the phase-conjugate mode.

\acknowledgements
This work is supported in part by the U.S. Office of Naval Research
Contract No. 14-91-J1205, by the National Science Foundation Grant
PHY95-07639, by the U.S. Army Research Office and by the
Joint Services Optics Program.


\end{document}